\documentclass[fleqn,10pt]{wlscirep}
\usepackage{graphicx}
\usepackage{subfigure}
\usepackage{mathrsfs}

\title{L\'{e}vy noise-induced transitions in gene regulatory networks}

\author[1,2]{Fengyan Wu}
\author[1,2,*]{Xiaoli Chen}
\author[1,2]{Yayun Zheng}
\author[3]{Jinqiao Duan}
\author[4,5]{J\"{u}rgen Kurths}
\author[3]{Xiaofan Li}

\affil[1]{Center for Mathematical Sciences, Huazhong University of Science and Technology, Wuhan 430074, China}
\affil[2]{School of Mathematics and Statistics, Huazhong University of Science and Technology, Wuhan 430074, China}
\affil[3]{Department of Applied Mathematics, Illinois Institute of Technology, Chicago, IL 60616, USA}
\affil[4]{Research Domain on Transdisciplinary Concepts and Methods, Potsdam Institute for Climate Impact Research, PO Box 60 12 03, 14412 Potsdam, Germany.}
\affil[5]{Department of Physics, Humboldt University of Berlin, Newtonstrate 15, 12489 Berlin, Germany}
\affil[*]{The corresponding author: xlchen@hust.edu.cn}

\begin{abstract}
Important effects of noise on a one-dimensional gene expression model involving a single gene have recently been discussed. However, few works have been devoted to the transition in two-dimensional models which include the interaction of genes. Therefore, we investigate here, a quantitative two-dimensional model (MeKS network) of gene expression dynamics describing the competence development in the B. subtilis under the influence of L\'evy as well as Brownian motions, where noises can do the B. subtilis a favor in nutrient depletion. To analyze the transitions between the vegetative and the competence regions therein, two deterministic quantities, the mean first exit time (MFET) and the first escape probability (FEP) from a microscopic perspective, as well as their averaged versions from a macroscopic perspective, are applied. The relative contribution factor (RCF), the ratio of non-Gaussian and Gaussian noise strengths, is adopted to implement optimal control in these transitions. Schematic representations indicate that there exists an optimum choice that makes the transition occurring at the highest probability. Additionally, we use a geometric concept, the stochastic basin of attraction, to exhibit a pictorial comprehension about the influence of the L\'{e}vy motion on the basin stability of the competence state.
\end{abstract}

\begin{document}
\flushbottom
\maketitle

\section*{Introduction}

Random fluctuations are ubiquitous in real life, and may change the fundamental dynamical properties of a system. A number of studies have reviewed that noises have rather counterintuitive and constructive effects on the behavior of a dynamical system.
Numerous research in different fields has demonstrated that noise can play a pivotal role in dynamical systems in biology,
physics, geophysics, chemistry, finance or engineering. This topic  has attracted more and more attention in recent years
 \cite{oksendal,Applebaum,DuanBook,Bressloff,Gardiner,Kampen}. The present contribution is devoted to investigating the influence of a combination of Gaussian and non-Gaussian noises on a gene regulatory network, where the interaction of genes is included in substantial extensions to recent work.

 It is known that messenger RNA (mRNA) and protein molecules are synthesized during the process of transcription and translation, respectively, which constitutes major procedures of gene expression. Bioscientists discovered that key transcription and regulatory factors pulse on and off repeatedly, often stochastically, even when cells are maintained in constant conditions. This type of spontaneous dynamic behavior is pervasive, appearing in diverse cell types from microbes to mammalian cells. The initiation of transcription, i.e., the binding of RNA polymerase at the promoter region of DNA, is inherently stochastic in biochemical reactions. Furthermore, the variability in the number of regulatory molecules provides noise sources from the extrinsic environment. Together, the inherent uncertainty in gene expression dynamics and external noisy environment make up the intrinsic and extrinsic noise sources \cite{Kaern05,Raj08,Eldar,Pal13,YangTao13}. Given that, it is necessary to take into account stochastic fluctuations when considering the dynamics of a gene expression system. It is often assumed that the noise is Gaussian. This arises due to the assumption that the perturbation is the result of a large number of independent interactions of bounded strength. However, this assumption is often not appropriate, in particular when the fluctuations are abrupt pulses or extreme events. Just as Levine et al \cite{Levine} pointed out that, pulsatile dynamics can produce ``long-tailed" distributions in static measurements based on flow cytometry and microscopy snapshots. Pulsatile dynamics is emerging as a central, and still largely unexplored area in the cell. In this situation, it is more appropriate to model the fluctuations by a process with heavy tails and discontinuous sample paths \cite{Applebaum,DuanBook,yayun16cf38}. L\'{e}vy motion is an appropriate prototype for non-Gaussian processes with jumps. Therefore, we will here study such a basic gene regulatory network under the influence of L\'{e}vy motion.

\begin{figure}[!t]
\centering
\includegraphics[width=\linewidth]{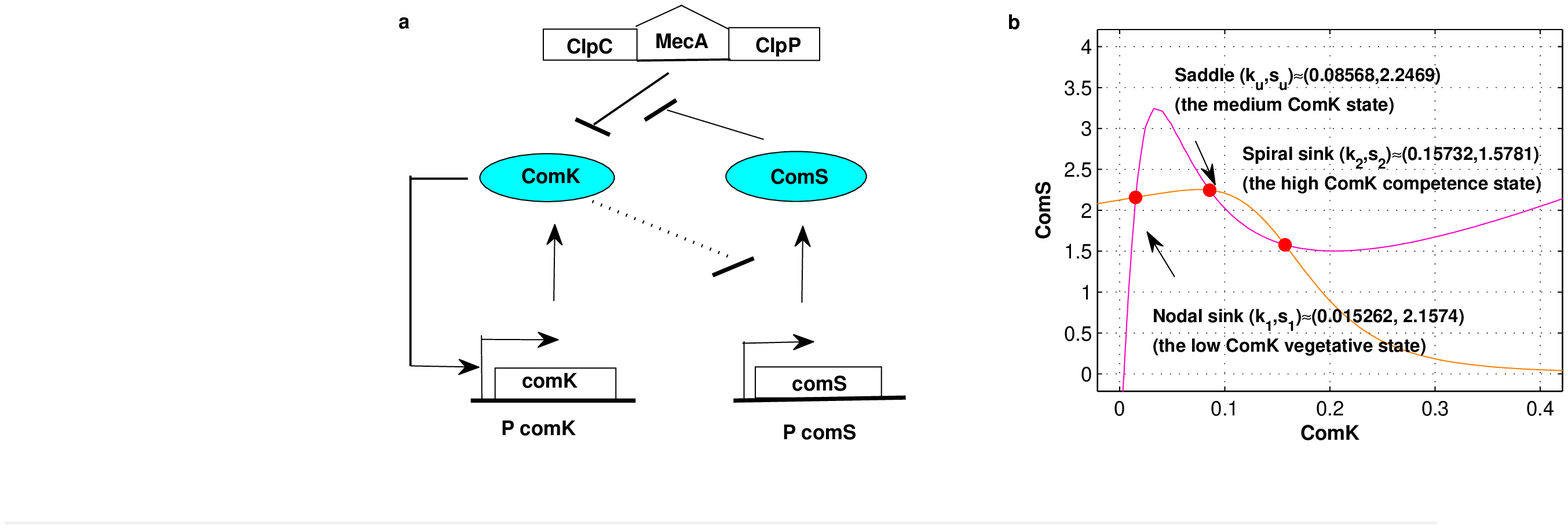}
\caption{\textbf{The core competence circuit in B. subtilis and nullclines for the MeKS model.} (a) $P_{comK} (P_{comS})$ denotes the promoter of gene $comK (comS)$. The arrow signifies the promotion relationship, while the short segment labels the inhibition relationship. ComK protein can active its own expression, forming a positive feedback loop.  Two ingredients, ComS protein competitively inhibits the degradation activity of the MecA-ClpP-ClpC complex which targets ComK protein and the indirect repression of comS gene by ComK protein, responsible for the negative feedback loop. (b) The intersection points are equilibria corresponding to nodal sink, saddle and spiral sink.}
\label{f1}
\end{figure}

Most of the B. subtilis population is in the vegetative state in which the comK gene expresses ComK protein at a low level. Under nutrient limitation, only a small fraction of B. subtilis transiently differentiates into the competence state in which the ComK protein at a high level, permitting the cell to take up extracellular DNA and to incorporate it into their own chromosome \cite{Dubnau99,Grossman95,Suel09,Suel15}, which is regulated by the MeKS network \cite{Suel06}. The core components of the MeKS network are shown in Fig.\ref{f1}a: the autoregulatory positive feedback loop regulated by the ComK protein, the ComS protein's competitive interference with the proteolytic degradation of the ComK protein, and the indirect repression of the comS gene by the ComK protein \cite{Dubnau98,Ogura99,Suel06}. The ComK protein can activate its own expression, shaping a positive feedback loop \cite{Dubnau05,Smits05}. Sinderen \cite{Sinderen} proposed that, the key transcription factor ComK protein, functioning as a master regulator, can activate the transcription of several genes which is necessary for the competence. When the concentration of the ComK protein is high, the competence develops \cite{Pal13cf12,Pal13cf20}, from which we know that the ComK protein occupies a central position in the competence-signal-transduction network.
The relationship between the ComK and the ComS protein is responsible for the formation of a negative feedback loop. Together, positive and negative feedback loops interacting with each other constitute the MeKS network.

In order to examine the dynamics of competence development governed by the MeKS network structure, basing on experimental observations, S\"{u}el et al \cite{Suel06} built a mathematical model \eqref{model} given in Methods, which involves both the direct positive and the ComS-related negative feedback loops of ComK. In the MeKS model \eqref{model}, the symbols $k$ and $s$ denote the concentration levels of the ComK and ComS proteins, respectively. The basal and fully activated rates of ComK production are $a_k$ and $b_k$, respectively. The parameter $k_0$ represents the concentration of the ComK protein needed for a $50\%$ activation. The  cooperativities of ComK auto-activation and ComS repression are indicated by the Hill coefficients $n$ and $p$, respectively. The ComS protein has the maximum expression rate $b_s$ and the half-maximal rate at $k=k_1$. The ComS protein competitively inhibits the degradation activity of the MecA-ClpP-ClpC complex that targets the ComK protein, which is responsible for the nonlinear degradation terms described at the end of the MeKS model \eqref{model}.

To understand the dynamics of the MeKS network without noise, with parameters suggested in ref.18, we calculate the nullclines in the ComK-ComS phase plane. Fig.\ref{f1}b exhibits three equilibria: two stable equilibria (nodal sink and spiral sink), whose ComK concentrations are low (vegetative state) and high (competence state), and an unstable saddle at a medium-concentration of ComK. The phase diagram of the MeKS model in the $b_k$-$b_s$ plane is depicted in refs.9 and 18
, which delineates four regions: (i) a monostability region in which only one stable equilibrium exists. (ii) a bistability region with three equilibria, two of them are stable. (iii) an excitability region with three equilibria ,only one of them is stable. (iv) a limit cycle oscillations region with only one unstable equilibrium. Our focus, however, is on the problem for the transition between both stable states. Therefore, the bistability region (ii) is chosen for our study.

It was found that stochasticity in gene expression can induce random transitions between both stable steady states, giving rise to binary distributions in the cell population, i.e., two pronounced subpopulations with different protein concentrations appear, respectively \cite{Kaern05}. Indeed, the influence of noise can cause random perturbations to the stable steady state of a system. Fluctuations  sufficiently strong enable the system to escape it, residing in another stable steady state \cite{Pal13,YangTao13}. The existence of two major ingredients, the positive feedback loop involved in dynamical system and the additional nonlinearity, contribute to the phenomenon of bistability \cite{Pal13cf10,Pal13cf11,Pal13cf12}.

According to previous works \cite{Pal13,Levine,YangTao14,yayun16,yayun16cf20}, noise is pervasive in the gene expression events. On occasion, these events show burst-like dynamical behaviors. Thus, Brownian motion and $\alpha$-stable L\'{e}vy motion is suitable to model the noise sources in gene regulatory systems. Here, the $\alpha$-stable L\'{e}vy motion has jumps, which is a special kind of non-Gaussian process defined by stable L\'{e}vy random variables. It has a larger jump magnitude with lower jump frequency when $\alpha$ closes to $0$, while smaller jump magnitude with higher jump frequency when $\alpha\in(1,2).$ It is substantially different from  Brownian motion whose paths are continuous but almost surely nowhere differentiable. The $\alpha$-stable L\'{e}vy motion's paths are continuous from the right and have left
limits (c\`{a}dl\`{a}g) at each time \cite{Applebaum,DuanBook}. Notice that a c\`{a}dl\`{a}g function can only have a countable number of jumps at most, permitting the system to transit between one stable state and another one which is enclosed in a non-adjacent domain.

\begin{figure}[!t]
\hspace{-0.5cm}
\includegraphics[width=7in]{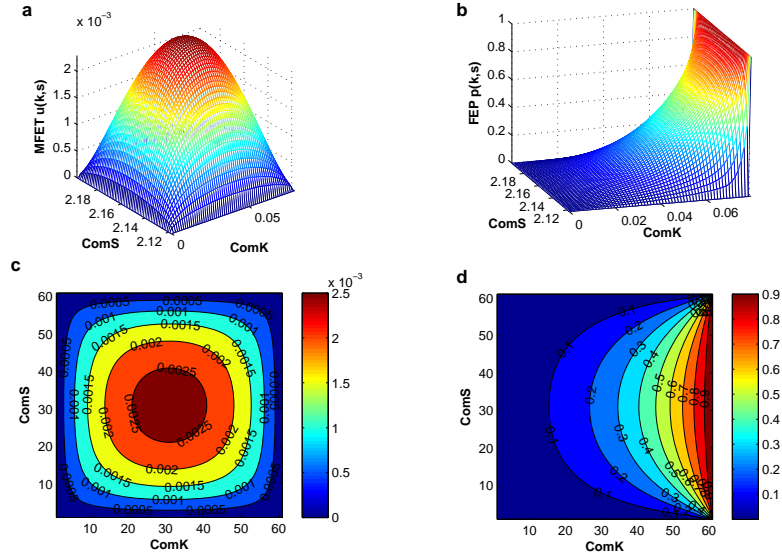}
\caption{\textbf{Pictorial representations of the MFET and the FEP in the two-dimensional stochastic MeKS model.} (a) MFET in the vegetative region $D\!=\!(0,0.0793)\!\times\!(2.11775,2.19705)$, when $\alpha=1.5,\sigma_k=\sigma_s=\varepsilon_k=\varepsilon_s=0.5.$ (b) FEP from the same vegetative region $D$ as in (a) to the adjacent competence region
$E\!=\!(0.0793,\infty)\!\times\!(2.11775,2.19705)$, when $\alpha=1.5,\sigma_k=\sigma_s=\varepsilon_k=\varepsilon_s=0.5.$ (c) Contour plot of the MFET for (a). (d) Contour plot of the FEP for (b).}
\label{f2}
\end{figure}

To examine the noise-induced transition between both stable states, two deterministic quantities, the mean first exit time (MFET) and the first escape probability (FEP) from a microscopic perspective, as well as the average first exit time (AMFET) and the average first escape probability (AFEP) from a macroscopic perspective, are utilised here. They are governed by differential-integral equations \eqref{mfet} and \eqref{fep} (see Methods), and a visualized representation for the MFET and the FEP in the two-dimensional stochastic MeKS model \eqref{stomodel} are exhibited in Fig.\ref{f2}. The microscopic perspective indicates that we analyze the dynamical behavior of a system from a certain point $(k,s)$, but in the macroscopic perspective the entirety dynamical behavior in a domain $D$ is regarded. Effects of L\'{e}vy and Brownian motions on a one-dimensional gene expression model have recently been investigated \cite{yayun16}. Here, a more general two-dimensional MeKS model describing the competence development in the B. subtilis under the influence of L\'evy as well as Brownian motion is studied. However, it becomes more challenging but difficult to conduct such a simulation in the two-dimensional model with L\'{e}vy motions. Nevertheless, we manage to calculate these deterministic quantities by accurate and efficient numerical methods \cite{yayun16cf37,yayun16cf38,yayun16cf39,li15,Xwang,lee}, instead of Monte-Carlo simulation.

The effects of noise on the model of autoregulatory gene expression involving a single gene has been discussed in ref.28.
However, the aforementioned work is only for a single gene activating a positive feedback loop, but avoids any consideration with other genes. However, the interplay of different genes is crucial for real genomic systems. Pal et al \cite{Pal13} studied a one-dimensional and a two-dimensional model in gene expression dynamics in B. subtilis, respectively. They obtained some early signatures of regime shifts in gene expression dynamics for the one-dimensional case, such as critical slowing down (CSD), rising variance and a ratio of two mean first passage times, while for the two-dimensional model, they did not get as many results as for the one-dimensional case. So far, little work has been devoted to effects of noise on the transition for the two-dimensional model of the MeKS network in B. subtilis. Our main goal is to utilise the knowledge of stochastic dynamics \cite{DuanBook} to gain some deeper insight into effects of various noises on the MeKS network, and in particular we try to explore and offer clues on the connection between vegetative growth and competence in the MeKS network.

Furthermore, it has been recently discovered that any disease progression can be divided into a normal state, a pre-disease state, and a disease state \cite{Kellershohn,Kim}, similar to the low vegetative state, the medium concentration state, and the high competence state in our work. As we know, diverse diseases pose a tremendous threat to people's health, and it has become a major challenge in medicine and biology for cancer research. In complex diseases networks, Huang \cite{huangsui} pointed out the gene expression noise induces effects on the choices between the normal and the malignant phenotype of the cell during cancer progression. Also, Jia et al \cite{jiachen} investigated the stochastic phenotype switching within the bacterial populations, and proposed the importance of the identification of the critical state for an early diagnosis in complex diseases networks. However, they just considered the influence of the Gaussian noise. Here, through our exploration for transitions between both stable states in gene expression, we expect our work to provide clues for the prevention and therapy of complex diseases system with a normal state, a pre-disease state, and a disease state.

\section*{Results}

To illuminate our work, we choose parameters of the bistability region (ii) as listed in ref.18 for our numerical experiments, namely, $a_k=0.004, b_k=0.14, b_s=0.68, k_0=0.2, k_1=0.222, n=2, p=5.$ Then, the deterministic MeKS model \eqref{model} has two stable states at $(k_l,s_l)\approx(0.015262, 2.1574)$ (the low ComK vegetative state), $(k_h,s_h)\approx(0.15732, 1.5781)$ (the high ComK competence state) and the unstable state at $(k_u,s_u)\approx(0.08568,2.2469)$ (the medium ComK state). The effects of noise on the one-dimensional model involving a single gene has been investigated \cite{yayun16,YangTao14}, and it has been recently found that the noise intensity and the size of jumps related with the L\'{e}vy motion index $\alpha$ can take effect on the concentrations within a cell in gene expression \cite{yayun16}. However, little work has been committed to the transition for the two-dimensional model. Here, we study a more general situation, i.e., a quantitative two-dimensional model (MeKS network) of gene expression dynamics describing competence development in B. subtilis with noises. Next, we present our results on the effects of Gaussian Brownian motion vs. non-Gaussian $\alpha$-stable L\'{e}vy motion on the first exit time and the first escape probability (both from microscopic and macroscopic perspectives) as well as basin stability for the competence state in the two-dimensional MeKS model.

\medskip
\noindent\textbf{Shorter MFET for higher noise intensity and smaller jump magnitude.}
For consideration of the residence time of the vegetative stable state, we fix the low vegetative region $D\!=\!(0,0.793)\!\times\!(0,2.46)$ enclosing the low ComK stable state $(k_l,s_l)$ from the ComK-ComS phase plane. Now, from microscopic and macroscopic perspectives respectively, we will investigate the influence of Brownian motion and $\alpha$-stable L\'{e}vy motion on the first exit time, which is the mean time of the ComK protein residing in the low vegetative region $D$ before exiting to the high competence region. In Methods, the differential-integral equations \eqref{mfet} offers how to get the mean first exit time $u(k,s)$, which is subject to the nonlocal ``Dirichlet'' condition, i.e. $u(k,s)$ vanishes outside of $D$.

Without stochastic perturbations, trajectories starting from the initial state around one stable state can not switch to the region enclosing another stable state. Thus, the residence time staying in the initial region is infinite. However, due to the influence of noise, the system can perform a given transition between both stable states.

 Here, we examine the impact of the Brownian motion and $\alpha$-stable L\'{e}vy motion on the mean first exit time, separately. For convenience, we introduce a notation $\Psi$ as the noise intensity under different stochastic fluctuations, i.e.,
\begin{eqnarray*}\label{noiseintensity}
\hspace{2.3cm}
  \left \{
  \begin{array}{ll}
     \Psi=\varepsilon_k=\varepsilon_s, \sigma_k=\sigma_s=0.  & \hbox{$\alpha\in(0,2)~~ (\alpha-stable~ L\acute{e}vy ~motion)$,} \\
     \Psi=\sigma_k=\sigma_s, \varepsilon_k=\varepsilon_s=0. & \hbox{$\alpha=2 ~~(Brownian ~motion)$.}
  \end{array}
\right.
\end{eqnarray*}

\begin{figure}[!t]
\hspace{-2cm}\includegraphics[width=9in]{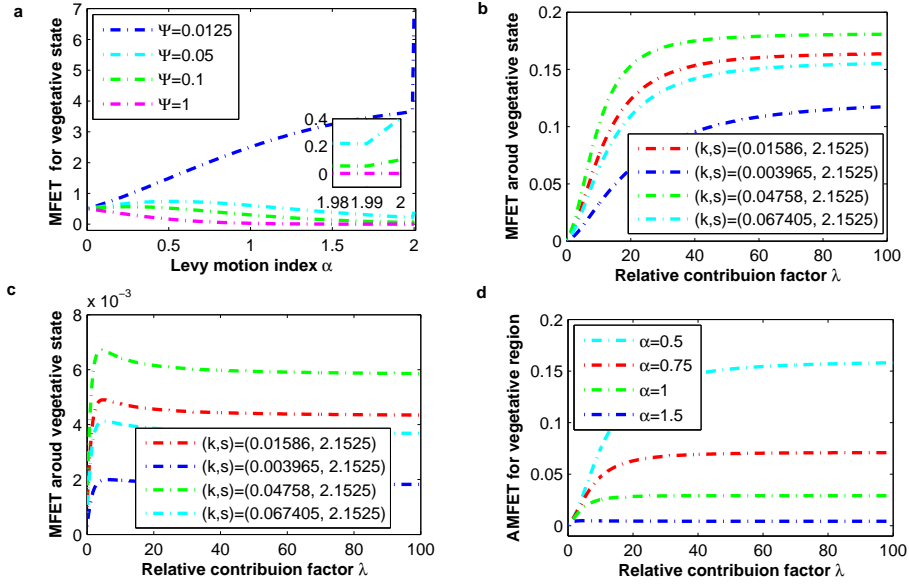}
\caption{\textbf{Effects of $\Psi$, $\alpha$ and $\lambda$ on the MFET for the low ComK vegetative region.} (a) MFET for the vegetative state (nodal sink) versus $\alpha$ with different noise intensities $\Psi$. (b) MFET of different initial concentrations versus $\lambda$, when $\alpha=0.5$. (c) MFET of different initial concentrations versus $\lambda$, when $\alpha=1.5$. (d) AMFET versus $\lambda$ for different $\alpha.$}
\label{f3}
\end{figure}

First, we analyse the noise intensity and L\'{e}vy motion parameter from a microscopic viewpoint (MFET for the low ComK vegetative state). It is shown in Fig.\ref{f3}a that, when the noise intensity is small enough, MFET increases as $\alpha$ increases, and grows rapidly at $\alpha=2$ (special case corresponding to Brownian motion). That is, if the noise intensity is slight, then the vegetative state is more stable under Brownian motion than $\alpha$-stable L\'{e}vy motion. Nonetheless, when the noise intensity becomes larger, the tendency of MFET changes from increase to decrease as $\alpha$ increases, implying that there are critical values of $\alpha$ resulting in a change of MFET from increase to decrease. Meanwhile, when $\alpha$ is fixed at a certain value, there is a decline in  MFET as the noise intensity $\Psi$ increases. In other words, $\Psi$ weakens the stability of the vegetative state. The macroscopic situation (AMFET for the low ComK vegetative region) is similar to the consequence in the microscopic situation, we omit its figure here. As shown in Fig.\ref{f3}a, a larger $\Psi$ induces a shorter residence time. In addition, compared with Brownian motion, the $\alpha$-stable L\'{e}vy motion makes MFET shorter when $\Psi$ are the same. Together, both microscopic and macroscopic situations demonstrate the consistency of the stability behaviour of the state $(k_l,s_l).$

The pictorial representation of the aforementioned results enlighten us on the transition subject, i.e., if we expect the ComK protein to exit from the low vegetative state to the high competence state, then a higher noise intensity and a larger $\alpha$ (smaller jump magnitude with higher frequency) should be selected.

\medskip
\noindent\textbf{Optimal control on the stability of the vegetative state.}
These results point out different effects of Gaussian noise (Brownian motion) and non-Gaussian noise ($\alpha$-stable L\'{e}vy motion) on the first exit time, individually. Now, we would like to discuss the influence of a combination of both types of noise on the first exit time. Due to the vital role of the ComK protein in the competence development, here, we apply the relative contributor factor (RCF) $\lambda$ proposed in ref.29.
In detail, $\lambda=\frac{\varepsilon_k}{\sigma_k}$, where $\sigma_k$ belongs to $(0,1]$, $\varepsilon_k$ stands for the non-Gaussian noise intensity and $\sigma_k$ for the Gaussian noise intensity, with the added constraints $\varepsilon_k=\varepsilon_s, \sigma_k=\sigma_s$ and $\varepsilon_k+\sigma_k=1$.

Fig. \ref{f3}b,c exhibit MFET around the low vegetative state (nodal sink) as a function of $\lambda$ for $\alpha=0.5$ and $\alpha=1.5$, respectively. From a microscopic viewpoint, we choose four different initial concentrations around the nodal sink, i.e., the red point is in close proximity to the nodal sink, the blue lies in left direction of the nodal sink, and the green and cyan in right direction of the nodal sink.
As shown in Fig.\ref{f3}b, with the increase of $\lambda$, all initial points of MFET increase fast, after reaching their maxima, the MFET tends to a constant. Meanwhile, the critical point $\lambda_0$ corresponding to the maximum of MFET is much larger than $1$, implying that the L\'{e}vy motion makes a much stronger contribution on MFET than the Brownian one. However, from Fig.\ref{f3}c, when $\alpha=1.5,$ the critical point $\lambda_0$ shifts towards a much smaller value than for $\alpha=0.5,$ signifying that the Brownian motion takes a more significant effect on the MFET gradually. Note that there is an obvious decline in MFET before arrival at a constant concentration. Furthermore, compared with Fig.\ref{f3}b,c, MFET for $\alpha=0.5$ is much higher for $\alpha=1.5.$

Next, we turn to the macroscopic viewpoint. The graphic representation of Fig.\ref{f3}d is more or less the same as Fig.\ref{f3}b. It is worth noticing that, when $\lambda$ is fixed, AMFET takes on a rising trend as $\alpha$ decreases, which implies that a smaller $\alpha$ can stabilize the vegetative state. Simultaneously, due to the  gradually significant effect on MFET by Brownian motion, the critical point $\lambda_0$ shifts towards a much smaller value as $\alpha$ increases.

Together, the critical point $\lambda_0$ corresponding to the maximum of MFET provides us the optimal control on the stability for the vegetative state. When the ratio of both noise intensities equals $\lambda_0$, it is more difficult to exit the vegetative region. In other words, noises determined by this ratio can enhance stability of the vegetative state. Actually, this is the phenomenon of noise- enhanced stability, a resonance-like effect, which has been found experimentally and theoretically in various systems\cite{YangTao14cf33,YangTao14cf34,YangTao14cf35,YangTao14cf36}. However, if we want to weaken the stability of the vegetative state, and expect the concentration of the ComK protein to acquire a high level, then a larger value of $\alpha$ is useful.

\medskip
\noindent\textbf{First escape probability.}
\begin{figure}[!t]
\hspace{-2.6cm}
\includegraphics[width=9in]{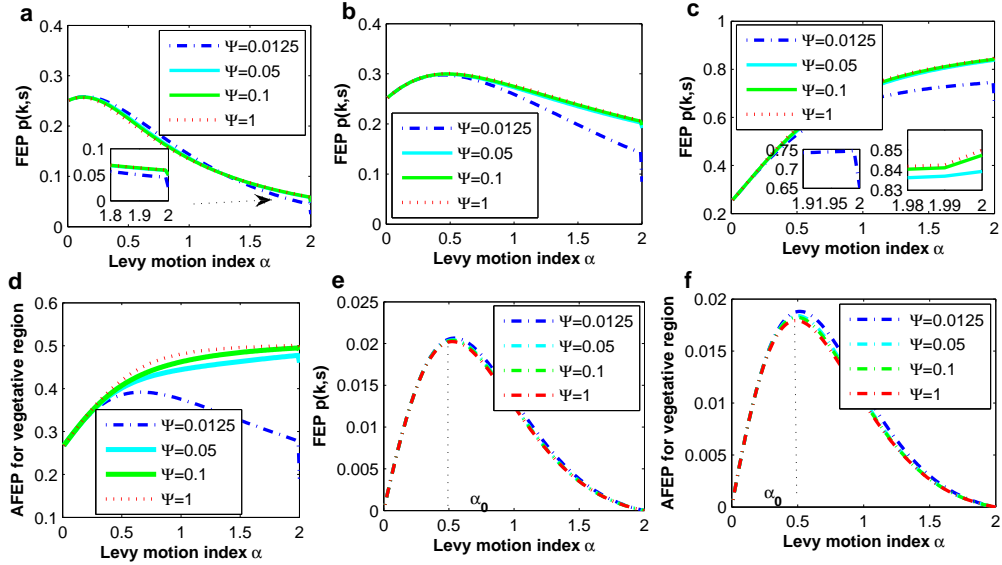}
\caption{\textbf{Behavior of escape probability from the vegetative region to the competence region for adjacent senario (a-d) and non-adjacent senario (e, f).} (a) Behavior of the FEP for the left point $(0.003965, 2.1525)$ of the vegetative state. (b) Behavior of the FEP for the point $(0.01586, 2.1525)$ in close to vegetative state (nodal sink $(0.015262, 2.1574)$).  (c) Behavior of the FEP for the right point $(0.067405, 2.1525)$ of the vegetative state. (d) Behavior of the AFEP for the vegetative region. (e) Behavior of the FEP for the vegetative state (nodal sink). (f) Behavior of the AFEP for the vegetative region.}
\label{f4}
\end{figure}
Now, we employ another concept to discuss the impacts of the noise intensity and $\alpha$ on the behavior of the transition probability, i.e., the likelihood of the first switching to a escape-target region. Here, we are concerned with the escape probability between the low vegetative region and the high competence region. According to the escaping phenomenon, there are two scenarios of the escape-target region: one is adjacent to the boundary of the starting region, but the other is non-adjacent. The first escape probability governed by the differential-integral equations \eqref{fep} can be solved by a numerical scheme efficiently in Methods.

\medskip
\noindent\textbf{Escape probability from the vegetative region to the adjacent competence region.}

In this part, we are concerned with the escape probability from the vegetative region $D\!=\!(0,0.0793)\!\times\!(0,2.46)$ to the adjacent competence region $E\!=\!(0.0793,\infty)\!\times\!(0,2.46)$. From a microscopic perspective, Fig.\ref{f4}a,b,c depict FEP as a function of $\alpha$ with different noise intensities. Fig.\ref{f4}a describes the behavior of the escape probability of the left point of the nodal sink. As shown in the figure, for different noise intensities, as $\alpha$ increases, there is an increasing tendency in FEP initially, but after reaching its maximum, FEP declines slowly. Notice that FEP is smallest at $\alpha=2$ (Brownian motion). From Fig.\ref{f4}b, we find that the behavior of the FEP of the point near the nodal sink is similar, except the much slower decline rate. However, for the right point of the nodal sink (Fig.\ref{f4}c), the FEP increases as $\alpha$ increases. Interestingly, due to its relative high concentration, the smallest noise intensity ($\Psi=0.0125$) makes FEP decreasing at $\alpha=2$, however, other noise intensities lead FEP to increase. Compared with Fig.\ref{f4}a,b,c, we observe that the transition probability varies with different initial concentrations, and it increases when the ComK protein initial concentration $(k,s)$ is close to the relative high concentration.

Now, we turn to look at the whole escape behavior. It is shown in Fig.\ref{f4}d that, for the smallest noise intensity ($\Psi=0.0125$), the escape behavior is about the same as the point near the nodal sink in Fig.\ref{f4}b. While for other noise intensities, AFEP increases as $\alpha$ increases, the AFEP drops at $\alpha=2$ (Brownian motion).

\medskip
\noindent\textbf{Escape probability from the vegetative region to the non-adjacent competence region.}

 The escape probability from the vegetative region $D\!=\!(0,0.04)\!\times\!(1.26,2.46)$ to the non-adjacent competence region $E\!=\!(0.138,0.178)\!\times\!(1.26,2.46)$ including the spiral sink is considered in this part, and the areas of the two regions are same. There exists a critical value of $\alpha_0$ with the highest escape probability in Fig.\ref{f4}e,f. First, from a microscopic perspective, Fig.\ref{f4}e depicts FEP as a function of $\alpha$ with different noise intensities. As $\alpha$ increases, there is an increasing tendency in FEP initially, but after reaching its maximum, FEP drops down to zero. Notice that FEP is much smaller than the adjacent scenario, and it is zero at $\alpha=2$, due to the fact that Brownian motion's trajectories are continuous, thus they can not jump. This last result is in accordance with the existing work\cite{yayun16cf32,yayun16cf33,yayun16}. Macroscopically, as shown in Fig.\ref{f4}f, the escape behavior of the whole vegetative region enclosing the nodal sink is similar.

 \medskip
\noindent\textbf{Optimal control for FEP from the vegetative region to the adjacent competence region.}
\begin{figure}[!t]
\hspace{-2.6cm}
\includegraphics[width=9in]{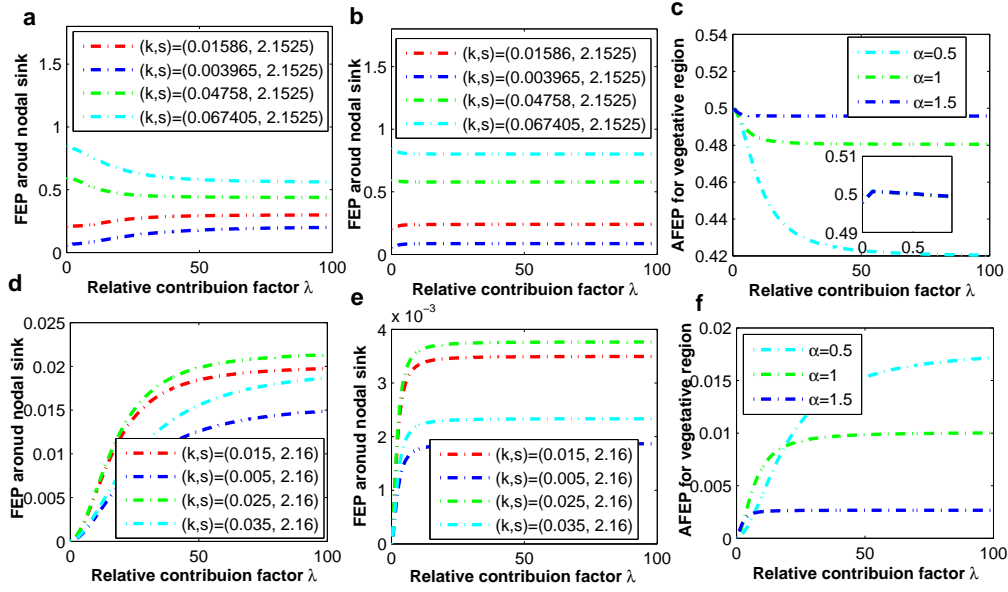}
\caption{\textbf{Effects of RCF $\lambda$ on the first escape probability for adjacent senario (a-c) and non-adjacent senario (d-f)}. (a) FEP of different initial concentrations versus RCF, when $\alpha=0.5$. (b) FEP of different initial concentrations versus RCF, when $\alpha=1.5$. (c) AFEP versus RCF with different $\alpha.$ (d) FEP of different initial concentrations versus RCF, when $\alpha=0.5$. (e) FEP of different initial concentrations versus RCF, when $\alpha=1.5$. (f) AFEP versus RCF with different $\alpha.$}
\label{f5}
\end{figure}

To implement the optimal control for the first escape probability of the transition from the vegetative region $D\!=\!(0,0.0793)\!\times\!(0,2.46)$ to the adjacent competence region $E\!=\!(0.0793,\infty)\!\times\!(0,2.46)$, also, the relative contributor factor $\lambda$ is utilised. Microscopically,  in Fig.\ref{f5}a,b, the same four initial concentrations around the low vegetative state(nodal sink) in Fig.\ref{f3}b,c are selected. As shown in Fig.\ref{f5}a, when $\alpha=0.5$, before entering to their constant concentrations, there is an opposite tendency for different points with the increase of $\lambda$, i.e., FEP is decreasing for the two right points of the nodal sink, while the point near the nodal sink and the left point of the nodal sink increase. Meanwhile, the critical value $\lambda_0$ corresponding to minima or maxima is large for $\alpha=0.5$, which indicates that the L\'{e}vy motion brings much more contribution to the transition. Then, when $\alpha=1.5$, there is a similar tendency in Fig.\ref{f5}b with Fig.\ref{f5}a, except for the position of the critical value $\lambda_0$.
Now, macroscopically, in Fig.\ref{f5}c, with the increase of $\lambda$, a slight increase in AFEP emerges for a much smaller $\lambda$, and after reaching its maximum, AFEP declines to arrive at a constant. When $\lambda$ is fixed at some value, AFEP increases as $\alpha$ increases. This indicates that we can choose a larger $\alpha$ (a smaller jump magnitude with a higher frequency) to make the transition occur at the highest probability.

\medskip
\noindent\textbf{Optimal control for FEP from the vegetative region to the non-adjacent competence region.}

Following up with the optimal control for the adjacent scenario, the relative contributor factor $\lambda$ is utilised too, for the  transition from the vegetative region $D\!=\!(0,0.04)\!\times\!(1.26,2.46)$ to the non-adjacent competence region $E\!=\!(0.138,0.178)\!\times\!(1.26,2.46)$. Microscopically, four initial concentrations around the low vegetative state (nodal sink) are selected. It is shown in Fig.\ref{f5}d, when $\alpha=0.5$, that all the selected points increase with the increase of $\lambda$. Meanwhile, the critical value $\lambda_0$ corresponding to the maximum is large for $\alpha=0.5$, implying that the L\'{e}vy motion gives much more contribution to the transition. When $\alpha=1.5$, the tendency in Fig.\ref{f5}e is the same with Fig.\ref{f5}d, but the position of the critical value $\lambda_0$ shifts towards a much smaller point. Macroscopically, as shown in figure Fig.\ref{f5}f, AFEP increases as  $\lambda$ increases. When $\lambda$ is fixed at a certain value, AFEP increases as $\alpha$ decreases. Hence, the results indicate that $\alpha_0\approx0.5$ (a larger jump magnitude with a lower frequency) is an ideal choice to make the transition to a high level at the highest probability.

\medskip
\noindent\textbf{FEP between the vegetative region and the competence region for different scenarios.}

\begin{figure}[!t]
\hspace{-0.5cm}
\includegraphics[width=7in]{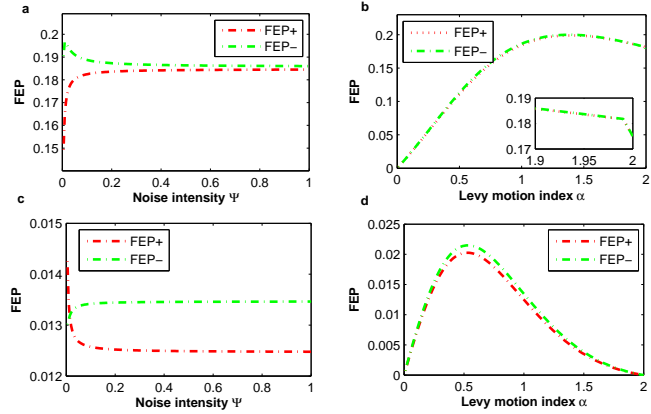}
\caption{\textbf{Transition between the vegetative region and the competence region for different scenarios.} (a) Behavior of the FEP between vegetative region and the adjacent competence region, when $\alpha=1.$ (b) Behavior of the FEP between vegetative region and the adjacent competence region, when $\sigma_k=\sigma_s=\varepsilon_k=\varepsilon_s=0.5.$ (c) Behavior of the FEP between vegetative region and the non-adjacent competence region, when $\alpha=1.$ (d) Behavior of the FEP between vegetative region and the non-adjacent competence region, when $\sigma_k=\sigma_s=\varepsilon_k=\varepsilon_s=0.5.$ }
\label{f6}
\end{figure}

The preceding results exhibit the behavior of escape probability in one direction, i.e., from the vegetative region to the competence region. S\"{u}el et al \cite{Suel06} pointed that individual cells in B. subtilis can switch to an alternative state and, after some time, switch back again. It is necessary to survey the behavior of escape probability in both directions. We take notations FEP+ as the FEP from the vegetative state (nodal sink) to the competence region, and FEP- the FEP from the competence state (spiral sink) to the vegetative region. Fig.\ref{f6} describes the behavior of the FEP in both directions for different scenarios from a microscopic perspective. Two regions with equal areas, the low vegetative region $D_1\!=\!(0,0.086291)\!\times\!(1.26,2.46)$ and the adjacent high competence region $\tilde{D}_1\!=\!(0.086291,0.172582)\!\times(1.26,2.46)\!,$ are selected in Fig.\ref{f6}a,b. And for the non-adjacent scenario (Fig.\ref{f6}c,d), we choose regions $D_2\!=\!(0,0.04)\!\times\!(1.26,2.46)$ and $\tilde{D}_2\!=\!(0.138,0.178)\!\times\!(1.26,2.46)$.

Fig.\ref{f6}a shows the behaviors of FEP in both directions when $\alpha=1.$ We find that FEP- is higher than FEP+, implying the transition from the competence state (spiral sink) to the adjacent vegetative region is more likely to occur. Then, when all the noise intensity are fixed at $0.5$ in Fig.\ref{f6}b, with the increase of $\alpha,$ FEP in both directions initially increase, after reaching their maxima, decreases gradually. As shown in Fig.\ref{f6}c, when $\alpha=1,$ in particular with the increase of the noise intensity, FEP- increases, but FEP+ decreases. Then, when all the noise intensity are fixed at $0.5$ in Fig.\ref{f6}d, with the increase of $\alpha,$ FEP in both directions initially increase, after reaching their maxima, drops to $0$ gradually. Macroscopically, The behavior of AFEP in both directions are almost same for different scenarios, we omit here.

From the proceeding results on the transition between the vegetative and the competence region, when noise intensity is slightly larger, no matter what escape scenarios we choose, the transition from the high competence region to the low vegetative region is more likely to happen. In the following we explain how to reduce this probability, and how to get the system to reside a longer time in the competence region.

\medskip
\noindent\textbf{Stochastic basin of attraction}

 We use now the concept of the basin of attraction of an invariant set \cite{Meiss}, i.e., the set whose solution of the system starting from an initial state in it ultimately converges to it. Inspired by refs.29 and 47,
we would like to extend the concept of stochastic basin of attraction (SBA), to make a quantification for the stability of a given competence state(spiral sink) in $\mathbb{R}^{2}$.

  \emph{The  stochastic basin of a closed set $K$ is denoted by $\mathfrak{B}(K) \triangleq  K \bigcup M$. We employ MFET $u(k,s)$ to quantify the closed set $K$,  $K = \{(k,s) \in B | u(k,s) \geq u^*\},$ in other words, the stochastic solution starting in  $K$ and stays there for a finite time, and $B$ is the selected deterministic basin of attraction of an invariant set. Thus, in the sense of stochastic dynamics,  the closed set $K$ can be taken as a stochastic invariant set. It is the collection of all initial states $(k_0,s_0)\in \mathbb{R}^{2}$ satisfying: $M = \{(k_0,s_0) \in K^{c} | p(k_0,s_0) > p^*\}$, where $p(k_0,s_0)$ represent the probability for all initial points returning to the closed set $K$ from outside of $K$.}

\begin{figure}[!t]
\hspace{-2.6cm}
\includegraphics[width=9in]{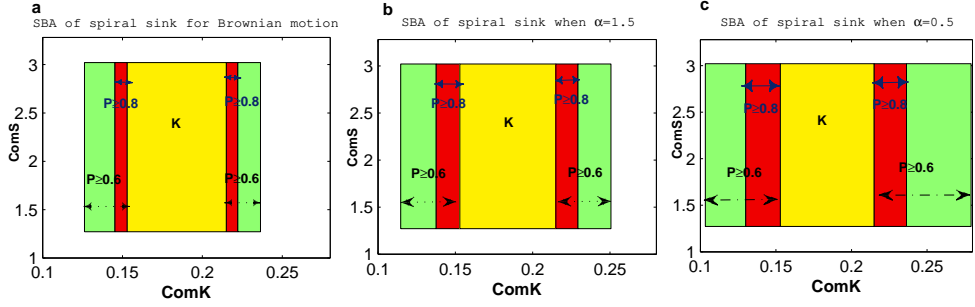}
\caption{\textbf{Impact of noises on the stochastic basin of attraction (SBA) of the competence state (spiral sink).} (a) SBA for competence state under Brownian motion, $\sigma_k=\sigma_s=1, \varepsilon_k=\varepsilon_s=0$. (b) SBA for competence state under $\alpha$-stable L\'{e}vy motion with $\alpha=1.5,  ~\sigma_k=\sigma_s=\varepsilon_k=\varepsilon_s=0.5.$ (c) SBA for competence state under $\alpha$-stable L\'{e}vy motion with $\alpha=0.5,~\sigma_k=\sigma_s=\varepsilon_k=\varepsilon_s=0.5.$}
\label{f7}
\end{figure}

 Here we suppose that the total noise intensities are fixed, and select the set $B\!=\!(0.115,0.253)\!\times\!(1.06,3.18)$. According to MFET under three fluctuations as listed in Fig.\ref{f7}, we choose $u^*=0.0038$ and determine $K\!=\!(0.15295,0.21505)\!\times\!(1.06,3.18).$ The colorful region is the SBA for the competence state (spiral sink).  The yellow area stands for the stochastic invariant set $K$ in which the initial points will stay a longer time. Undoubtedly, the competence state (spiral sink) is included in $K.$ The set which has a probability with $p^*=0.6$ returns to $K$ is labeled in red and green colors, and the probability with $p^*=0.8$ in red color. Fig.\ref{f7} exhibits a pictorial comprehension for the basin stability of competence state, and there is an expansion in SBA with $\alpha$ decreases.  That is, the high competence state is more stable under fluctuations of Brownian motion and $\alpha$-stable L\'evy motion with a smaller $\alpha$. In other words, L\'{e}vy motion with a larger $\alpha$ is more likely to induce the transition to the low vegetative state.

\section*{Discussion}\label{conclusion}

To sum up, we have investigated the impacts of stochastic fluctuations on the evolutionary process of the MeKS network system, where (Gaussian) Brownian and (non-Gaussian) $\alpha$-stable L\'evy motion are both incorporated. Two deterministic quantities, the mean first exit time (MFET) and the first escape probability (FEP) from a microscopic perspective, the average first exit time (AMFET) and the average first escape probability (AFEP) from a macroscopic perspective, are employed to analyse the transitions between the vegetative and the competence region in the networks. As we have known that the effects of L\'{e}vy and Brownian motions on a one-dimensional gene expression model have recently been investigated \cite{yayun16}, we here study a more general two-dimensional MeKS model describing the competence development in the B. subtilis under the influence of L\'evy as well as Brownian motion. However, to conduct such a simulation in the two-dimensional model with L\'{e}vy motions is such a more challenging but difficult problem. Nevertheless, we have managed to calculate these deterministic quantities by accurate and efficient numerical methods \cite{yayun16cf37,yayun16cf38,yayun16cf39,li15,Xwang,lee}.

The noise intensities $\sigma_k, \sigma_s, \varepsilon_k,\varepsilon_s$, the L\'{e}vy motion index $\alpha$  (relating with the jump magnitude and its frequency) have important and subtle influences on MFET and FEP. We have used the relative contribution factor (RCF) $\lambda$, the ratio between non-Gaussian and Gaussian noise strengths, to implement optimal control in the transition between the two states, when the total quantities of the noise intensities are fixed. We here found that MFET as a function of $\lambda$ shows a maximum, signifying the emergence of the noise-enhanced stability phenomenon for the low vegetative state.

For different initial concentrations, the behavior of the escape probability from the vegetative region to the adjacent competence region is diverse, depending on the influences of the noise intensities and L\'{e}vy motion index $\alpha$. The $\alpha$-stable L\'evy motion can induce a transition between the vegetative region and a non-adjacent competence region, while Brownian motion can not implement it, due to its continuous paths. Pictorial representations of FEP versus RCF indicate that, for the adjacent scenario, a larger $\alpha$ (a smaller jump magnitude with a higher frequency) makes it happen at the highest probability. While for the non-adjacent scenario, $\alpha_0\approx0.5$ (a larger jump magnitude with a lower frequency) is the best choice. Also, we have surveyed the behavior of escape probability for the transition between the vegetative and the competence region with respect to the two escape scenarios (adjacent, non-adjacent).

Additionally, we have explored the stochastic basin of attraction for the competence state from a geometric viewpoint. Through a schematic presentation of the influence of the L\'{e}vy motion index $\alpha$ on the basin stability of the competence region in our stochastic MeKS model, we have detected that the high competence state is more stable under combined fluctuations of Brownian motion and $\alpha$-stable L\'evy motion with a smaller $\alpha$.

Furthermore, various diseases pose a tremendous threat to people's health. Cancer research has become a major challenge in medicine and biology. In complex diseases networks, the gene expression noise induces effects on the choices between the normal and the malignant phenotype of the cell during disease progression. Here, we hope that our exploration will offer a new guidance for the prevention and therapy of complex diseases system with a normal state, a pre-disease state and a disease state.

\section*{Methods}

\textbf{The dynamical model of competence induction.}
To understand how the MeKS network structure determines the dynamics of competence, S\"{u}el et al  \cite{Suel06} built a mathematical model constrained by experimental observations. This model can be reduced to a system of two ordinary differential equations incorporating both the direct positive and the ComS-regulated negative feedback loops of ComK.
\begin{equation}  \label{model}
 \begin{split}
 \hspace{5.8cm}
\tfrac{dk}{dt}&={{a}_{k}}+\tfrac{{{b}_{k}}{{k}^{n}}}{k_{0}^{n}+{{k}^{n}}}-\tfrac{k}{1+k+s},\\
 \tfrac{ds}{dt}&=\tfrac{{{b}_{s}}}{1+{{(k/{{k}_{1}})}^{p}}}-\tfrac{s}{1+k+s}.
 \end{split}
\end{equation}

 For convenience, we name it as MeKS model and write $f(k,s)={{a}_{k}}+\tfrac{{{b}_{k}}{{k}^{n}}}{k_{0}^{n}+{{k}^{n}}}-\tfrac{k}{1+k+s}$, and $g(k,s)=\tfrac{{{b}_{s}}}{1+{{(k/{{k}_{1}})}^{p}}}-\tfrac{s}{1+k+s}.$ In the MeKS model, the symbols $k$ and $s$ stand for the concentrations of ComK and ComS protein, respectively. $a_k$ and $b_k$ denote the basal and fully activated rates of ComK production, and $k_0$ the concentration of the ComK protein needed for $50\%$ activation. The cooperativities of ComK auto-activation and ComS repression are prescribed by the Hill coefficients $n$ and $p$, respectively. $b_s$ is the maximum expression rate of the ComS protein, and $k=k_1$  has the half-maximal rate. The ComS protein competitively inhibits the degradation activity between the MecA-ClpP-ClpC complex and the ComK protein, which leads to the nonlinear degradation terms described at the end of the MeKS model.

   It is seen from their nullclines in Fig.\ref{f1}b that there are three equilibria: Nodal sink $(k_l,s_l)\approx(0.015262, 2.1574)$ (the low stable ComK vegetative state), saddle $(k_u,s_u)\approx(0.08568,2.2469)$ (the unstable medium ComK state), and spiral sink $(k_h,s_h)\approx(0.15732, 1.5781)$ (the high stable ComK competence state).

\medskip
\noindent\textbf{The stochastic dynamical model of competence induction.}
To investigate effects of stochastic fluctuations on the transitions between the vegetative and the competence region, our stochastic MeKS model is described as
\begin{equation}  \label{stomodel}
 \begin{split}
 \hspace{5.3cm}
\tfrac{dk}{dt}&=f(k,s)+\sigma_k \dot{B}_t^1+\varepsilon_k \dot{L}_t^1, \\
 \tfrac{ds}{dt}&=g(k,s)+\sigma_s \dot{B}_t^2+\varepsilon_s \dot{L}_t^2.
 \end{split}
\end{equation}
Two kinds of noises, Gaussian and non-Gaussian noises, are incorporated in the MeKS model. The symbols $\sigma_k,\sigma_s$ represent the intensities of the Gaussian noise, and $\varepsilon_k, \varepsilon_s$  the intensities of the non-Gaussian noise. The symbols $B_t^1$ and $B_t^2$ are two independent Brownian motions, $L_t^1$ and $L_t^2$ are two independent $\alpha$-stable L\'{e}vy motions, which is
a pure jump process and independent of Brownian motions.  Due to the vital role of the ComK protein in the competence development, to implement the optimal control of noises for transitions between the vegetative and the competence region, we employ the relative contribution factor $\lambda$, which is the ratio of the intensity of Gaussian noise and non-Gaussian noise, i.e.,
 $\lambda=\frac{\varepsilon_k}{\sigma_k}$, where $\sigma_k$ belongs to $(0,1]$, with the added restrictions $\varepsilon_k=\varepsilon_s, \sigma_k=\sigma_s$ and $\varepsilon_k+\sigma_k=1$.

\medskip
\noindent\textbf{Brief introduction for L\'{e}vy motion.}
A L\'{e}vy motion $L_t$ is a stochastic process which has stationary and independent increments, and is regarded as an appropriate model for non-Gaussian processes with jumps and heavy tails \cite{sato}. In detail, for any time $t_1, t_2$ with $t_1<t_2,$ $L_{t_2}-L_{t_1}$ and $L_{t_2-t_1}$ have the same distribution. Namely, the distribution of $L_{t_2}-L_{t_1}$ only depends on the time difference $t_2-t_1.$ And for any partition $0=t_0<t_1<...<t_n=t, L_{t_i}-L_{t_{i-1}}$ are independent. The sample paths of L\'{e}vy motion are continuous from the right and have left limits at every time (c\`{a}dl\`{a}g)\cite{Applebaum,DuanBook}. L\'{e}vy motion, as a non-Gaussian stochastic process, is a generalization of Gaussian Brownian motion whose sample paths are continuous in time in the common sense.

In our present work, the $\alpha$-stable L\'{e}vy motion with a triplet $(0,0, \nu_\alpha)$, i.e., a pure jump motion is considered. The parameter $\alpha$ is called L\'{e}vy motion index, or stability index. Specially, it is a Brownian motion when $\alpha=2.$ For $0<\alpha<2,$ the jump measure in two-dimensional case is defined as
$\nu_{\alpha} = \frac{C_\alpha dy}{\|y\|^{2 + \alpha}},$
 where $C_\alpha=\frac{\alpha}{2^{1-\alpha}{\pi}}\frac{\Gamma(1+\frac{\alpha}{2})}{\Gamma(1-\frac{\alpha}{2})}$ . The $\alpha$-stable L\'{e}vy motion has a larger jump magnitude with a lower jump frequency for $0<\alpha<1$, while it has a smaller jump magnitude with higher jump frequencies for $\alpha$ closes to 2. For more details, we refer readers to an uncomplete list of references \cite{Applebaum,sato,DuanBook,sky} etc.

\medskip
\noindent\textbf{Mean first exit time (MFET) and average mean first exit time (AMFET).}
The MFET and the AMFET, are utilised to describe the stability for certain stable state of the stochastic MeKS model. First, for a bounded domain $D$ enclosing certain equilibrium, for instance the vegetative state, we define the first exit time as
$\tau_D(\omega) = \inf\{t \geq 0, X_{0}=x, ~X_{t} \in  D^c\},$
where $D^c$ denotes the complement set of $D$ in $\mathbb{R}^{2}$. Then the mean exit time is denoted as $u(x)  \triangleq   \mathbb{E}  \tau_D(\omega) \geq 0$. It is the mean residence time in certain domain $D$  before exiting to another regime and used to our analysis from microscopic perspective. According to ref.6, the MFET $u(k,s)$ in our model can be solved by  the following differential-integral equations with the Dirichlet exterior boundary condition
\begin{equation} \label{mfet}
 \begin{split}
\hspace{5cm}
  \mathscr{A}u(k,s) &= -1, \qquad  (k,s) \in D,\\
  u(k,s)  &= 0  ~~~\qquad  (k,s) \in D^c,
 \end{split}
\end{equation}
where $D=(a,b)\times (c,d)$ with real numbers $a,b,c,d$.  Here, due to the independent properties of the Brownian and L\'{e}vy motion, the jump measure in two-dimensional case is divided into the summation of two one-dimensional jump measure \cite{sunxu}. Then  the generator $\mathscr{A}$ is
\begin{eqnarray}\label{generator}
\hspace{2cm}
\mathscr{A} u(k,s)&=&f(k,s) u_k (k,s) + g(k,s)u_s (k,s) +\frac{{\sigma_k}^2}{2}u_{kk} + \frac{{\sigma_s}^2}{2}u_{ss}\notag\\
&+&\varepsilon_k^\alpha C_\alpha \int_{\mathbb{R}\backslash\{0\}} \frac{u(k\!+\!k',s)\!-\!u(k,s)}{|k'|^{1+\alpha}}dk'
\!+\!\varepsilon_s^\alpha C_\alpha \int_{\mathbb{R}\backslash\{0\}} \frac{u(k,s\!+\!s')\!-\!u(k,s)}{|s'|^{1+\alpha}}ds'.
\end{eqnarray}

  Secondly, the AMFET on $D$  is defined as $\bar{u}(D)= \frac{1}{|D|}\int\int_D u(k,s)dkds$ for our analysis from the macroscopic perspective \cite{DuanBook,lee}.
\medskip

\medskip
\noindent\textbf{ First escape probability (FEP) and average first escape probability (AFEP).}
The escape probability here are employed to characterize the likelihood of the transition between the low vegetative and the high competence region for the MeKS system. The FEP $p(k,s)$ can be solved by the following differential-integral equations with the Balayage-Dirichlet exterior boundary condition
\begin{equation} \label{fep}
 \begin{split}
 \hspace{5cm}
  \mathscr{A}p(k,s) &= 0, \qquad  (k,s) \in D,\\
  p(k,s)  &=
  \left \{
  \begin{array}{ll}
    1, & \hbox{$(k,s) \in E$;} \\
    0, & \hbox{$(k,s) \in D^{c}\setminus E$.}
  \end{array}
\right.
 \end{split}
\end{equation}
Here $ \mathscr{A}$ is the same generator as \eqref{generator}.

The AFEP on $D$  is defined as $\bar{p}(D)= \frac{1}{|D|}\int\int_D p(k,s)dkds$  \cite{DuanBook,lee}. We use the FEP to examine the transitions from a microscopic viewpoint, and the AFEP from the macroscopic perspective.

 It has recently been examined that the effects of L\'{e}vy and Brownian motions on a one-dimensional gene expression model  \cite{yayun16}. Here, we study a more general stochastic two-dimensional MeKS model describing the competence development in the B. subtilis. However, it turns into a more challenging but difficult problem to conduct such a simulation in the two-dimensional model with L\'{e}vy motions. Nevertheless, we have managed to apply recent results to calculate these deterministic quantities by accurate and efficient numerical methods \cite{yayun16cf37,yayun16cf38,yayun16cf39,li15,Xwang,lee}. Fig.\ref{f2} exhibits a visualized representation for the MFET and the FEP in the two-dimensional stochastic MeKS model \eqref{stomodel}.

\medskip
\noindent\textbf{Numerical methods.}\label{scheme}
To solve the MFET and the FEP efficiently and accurately, Gao et al \cite{yayun16cf37} constructed an effective numerical method to solve the aforementioned differential-integral equations.  They used the central difference
for derivatives and  a modified trapezoidal rule to discretize the differential-integral equations \cite{yayun16cf37}.
Their scheme can be modified  for our model with two transformations, i.e., $k=\frac{b-a}{2}z + \frac{b+a}{2}$ for $z \in (-1, 1)$, and  $s=\frac{d-c}{2}w + \frac{d+c}{2}$ for $w \in (-1, 1)$, correspondingly,  $k' = \frac{b-a}{2}\hat{k}$ and   $s' = \frac{d-c}{2}\hat{s}$. We let $v(z,w) = u(\frac{b-a}{2}z + \frac{b+a}{2},\frac{d-c}{2}w + \frac{d+c}{2})$, and address the integral term in equation \eqref{generator} by the Cauchy principal value integral. Then the generator  $ \mathscr{A}$ can be rewritten as
\begin{eqnarray*}\label{generator2}
\mathscr{A}\!v(z,\!w)&=&\frac{2}{b\!-\!a}\!f(\frac{b\!-\!a}{2}z \!\!+\!\! \frac{b\!+\!a}{2}\!,\!\frac{d\!-\!c}{2}w \!\!+ \! \!\frac{d\!+\!c}{2})v_z(z,w)\!\!+\!\!
\frac{2}{d\!-\!c}g(\frac{b\!-\!a}{2}z \!\!+\!\! \frac{b\!+\!a}{2},\frac{d\!-\!c}{2}w \!\!+\!\! \frac{d\!+\!c}{2})v_w(z,w)\\
&+&\frac{4}{{(b-a)}^2}\frac{\sigma_k^2}{2}v_{zz}+\frac{4}{(d-c)^2}\frac{\sigma_s^2}{2}v_{ww}-
\frac{\varepsilon_k^{\alpha}C_\alpha}{\alpha}(\frac{2}{b-a})^\alpha[\frac{1}{(1+z)^\alpha}+ \frac{1}{(1-z)^\alpha}]v(z,w)\\
&-&\frac{\varepsilon_s^{\alpha}C_\alpha}{\alpha}(\frac{2}{d\!-\!c})^\alpha[\frac{1}{(1\!+\!w)^\alpha}\!+\!\! \frac{1}{(1\!-\!w)^\alpha}]v(z,\!w)\!+\!
\varepsilon_k^\alpha C_\alpha (\frac{2}{d\!-\!c})^\alpha  \int_{\!-1-\!z}^{1\!-\!z} \frac{v(z\!+\!\hat{k},w)\!\!-\!\!v(z,w)}{|\hat{k}|^{1+\alpha}}d\hat{k}\\
&+&\varepsilon_s^\alpha C_\alpha (\frac{2}{d-c})^\alpha  \int_{-1-w}^{1-w} \frac{v(z,w+\hat{s})-v(z,w)}{|\hat{s}|^{1+\alpha}}d\hat{s}=\phi(z,w).
\end{eqnarray*}
Here $\phi(z,w)$ stands for the right-hand side term occurring at the differential-integral equation. For the equation \eqref{mfet}, $\phi(z,w)=-1.$ While for the equation \eqref{fep}, the expression of $\phi(z,w)$ varies with the different escape scenarios. In detail, first, for the adjacent scenario, if the system escapes from region $D=(a,b)\times(c,d)$ to an adjacent region  $E=(b,i)\times(c,d),$ then $\phi(z,w)=\frac{\varepsilon_k^{\alpha}C_\alpha}{\alpha}\{(i-\frac{b-a}{2}z - \frac{b+a}{2})^{-\alpha}-[\frac{b-a}{2}(1-z)]^{-\alpha}\},$ especially, for $i=\infty,$ $\phi(z,w)=-\frac{\varepsilon_k^{\alpha}C_\alpha}{\alpha}(\frac{2}{b-a})^\alpha\frac{1}{(1-z)^\alpha}.$ Reversely, if the system escapes from region $D=(a,b)\times(c,d)$ to an adjacent region  $E=(h,a)\times(c,d),$ then $\phi(z,w)=\frac{\varepsilon_k^{\alpha}C_\alpha}{\alpha}\{(\frac{b-a}{2}z + \frac{b+a}{2}-h)^{-\alpha}-[\frac{b-a}{2}(1+z)]^{-\alpha}\}.$ Secondly, for the non-adjacent scenario, if the system escapes from region $D=(a,b)\times(c,d)$ to a non-adjacent region  $E=(h,i)\times(c,d),$ then $\phi(z,w)=\frac{\varepsilon_k^{\alpha}C_\alpha}{\alpha}\{(i-\frac{b-a}{2}z - \frac{b+a}{2})^{-\alpha}-(h-\frac{b-a}{2}z - \frac{b+a}{2})^{-\alpha}\}.$ Reversely, if the system escapes from region $D=(a,b)\times(c,d)$ to a non-adjacent region  $E=(h,i)\times(c,d),$ then $\phi(z,w)=\frac{\varepsilon_k^{\alpha}C_\alpha}{\alpha}\{(\frac{b-a}{2}z - \frac{b+a}{2}-h)^{-\alpha}-(\frac{b-a}{2}z - \frac{b+a}{2}-i)^{-\alpha}\}.$

%

\section*{Acknowledgements}
We would like to thank Dongfang Li, Xiao Wang and Hansen Ha for discussions about computation. We are grateful to Aining Bai in Institute of Botany (CAS), Liang Wang, Mengli Hao, Jian Ren, Tao Jiang, Xinyong Zhang, Yanjie Zhang, Ziying He for helpful discussions. This work was partly supported by the NSFC grants 11531006, 11371367  and   11271290.

\section*{Author contributions statement}
F.W. and J.D. conceived the research and wrote the first draft of the manuscript. F.W. performed computations. X.C. and X.L. conducted and verified the program designs. Y.Z. and J.K. analysed the results and concepts development. All authors conducted research discussions and reviewed the manuscript.

\section*{Additional information}
\textbf{Competing financial interests:} The authors declare no competing financial interests.

\end{document}